# Magnetorheological Payne effect in bidisperse MR fluids containing Fe nanorods and $Fe_3O_4$ nanospheres: a dynamic rheological study

*Injamamul Arief, P.K. Mukhopadhyay[†]*

LCMP, Department of Condensed Matter Physics and Material Sciences, S. N. Bose National Centre for Basic Sciences, Salt Lake, Kolkata 700 098, India.

**Abstract:**

The spherical $Fe_3O_4$ with 300 nm in diameter was synthesized by typical thermal decomposition of Fe (III) organo-metallic precursor in polyol and polyacrylic acid. Fe-nanorods were prepared by reducing Fe (III) nitrate in presence of polyol-hydrazine-CTAB. Morphology and magnetic characterization of the nanoparticles were performed by ESEM, XRD and VSM studies. We performed detailed non-linear magnetorheological properties of three MR fluids (10 vol%) containing isotropic $Fe_3O_4$ and anisotropic Fe-nanorods under both small and large amplitude oscillatory flow. The MR samples demonstrated strong magnetorheological Payne effect i.e. rapid stress relaxation under increasing deformation and uniform magnetic field beyond linear viscoelastic region (LVR), which has not been studied in-depth in conventional MR fluids. We have also shown that stress softening was more pronounced for MR fluids with higher anisotropic contents, in contrast to isotropic MR fluid. The onset strains for LVR to non-linear region transition for anisotropic fluids were much lower than that of isotropic spherical nanoparticle-containing fluid. The stronger MR response for nanorod-containing MR fluids can be explained in terms of enhanced field-induced structuration.

Keywords: linear viscoelasticity, large amplitude oscillatory strain, bidisperse MR fluids, magnetorheological Payne effect, stress relaxation, field-induced structuration.

[†]Email: pkm@bose.res.in

# 1. Introduction

Magnetorheological fluids are comprised of magnetic microparticles in matrix fluids. The colossal change in magnetorheological properties of MR fluids can be explained in terms of formation of anisotropic microstructures upon application of magnetic field [1]. The magnetically polarized particles are connected to each other to form robust hierarchical chain structures, giving rise to field-induced viscosity and shear modulus. In order to form microstructure aggregates, hydrodynamic forces must be overcome by magnetostatic forces, therefore, factors influencing magnetic interactions such as magnetic properties, size and spatial distribution of magnetic particles also control MR effect [2]. Micron-sized spherical magnetic particles possess large application fields ranging from magnetic recording devices to biomedicines [3]. For many of these already documented applications, particle shape, size and magnetic saturation are the most important parameters that have to be considered. Particles with shape or structural anisotropy have been the subject of considerable interests in terms of their synthesis, chemical and physical properties etc. [4] Similar to isotropic spherical particles, these highly anisotropic nano and micro-structured materials can be used in many applications such as imaging and sensing. However, non-spherical particles e.g. rod-like microstructures can form stronger field-induced aggregates due to easy magnetization when the long axes are aligned parallel to magnetic field [5]. This gives rise to stronger MR response. Therefore, for a magnetorheological (MR) fluid to perform desirably, particle's surface morphology as well as shape anisotropy plays very crucial role.

An interesting trend in stress-strain behavior of rubber materials containing carbon black fillers was discovered back in 1962 and termed as Payne effect after the A. R. Payne [6]. This is manifested as the change in storage and loss modulus with increasing strain amplitude under constant frequency. At some critical amplitude, storage modulus tends to decrease rapidly while loss modulus shows maxima in the same region where G′ falls rapidly under deformation. Loss factor or tanδ (G″/G′) also increases rapidly in this region till it gets independent of strain. Although the effect was first demonstrated for rubber-polymer networks, similar trend was also reported recently for magnetic elastomers and polymer blends [7]. The term the "magnetic Payne effect" was first coined by An et al. for MR gels to highlight their strain-softening [8]. Recently, Sorokin et al. also reported magnetic Payne effect in carbonyl iron containing elastomer blends [7]. In both cases, polymer gels and elastomers were used as matrix for magnetic particles. Conventional MRFs with magnetized particles in mineral oils yet to have shown magnetorheological Payne effect. Therefore, detailed study of the effect in MRFs and ferrofluidic systems needs further attention. The focus of this paper is on magnetorheological Payne effect containing both isotropic and anisotropic particles in silicone oil suspensions.

One of the most common choices for magnetic microparticles in MR fluids is carbonyl iron particles (CIP) due to its high saturation magnetization. Magnetorheological studies in steady shear and oscillation modes have been reported with MRFs containing surface-modified monodispersed spherical CIP and magnetite microparticles [9]. However, tunable magnetic and surface properties of anisotropic rod-like microparticles have gained significant attention recently [10]. It has been reported that the particle settling in MRFs was markedly reduced when spherical particles were replaced by rod-like or fibrous materials [11]. Field-dependent solid frictions within MRFs also enhance MR properties. In this work, we reported the polyol-mediated synthesis of PAA-magnetite microspheres with excellent monodispersity and high chemical yield. A novel CTAB-mediated hydrothermal synthesis of iron microrods of average length of 700 nm was also described. Bidispersed MRFs prepared by substituting spherical magnetite particles with rod-like Fe-particles has shown enhanced magnetorheological Payne effect in oscillatory rheology, in contrast to MRF containing only spherical magnetite particles. In addition to superior magnetic properties of Fe-rods, shape anisotropy was also a key parameter that justifies the trends.

## 2. Experimental details

Spherical magnetite nanoparticles were synthesized by previously described thermal decomposition method of Fe (III) acetylacetonate (99%, Sigma-Aldrich) in liquid polyol. In this method, organometallic precursor solution in ethylene glycol (EG, Merck) was refluxed for 3 hours in presence of polyacrylic acid (PAA, avg. mol. wt. ~1800, Sigma-Aldrich). The molar ratio of polymer and Fe (III) acetylacetonate was maintained at 2:1. The appropriate amount of metal salt was stirred to dissolve in EG followed by rapidly heated to boil at $120^0$C for 30 minutes. Afterwards, the precursor was transferred to a round-bottom flask under steady nitrogen gas flow and refluxed at $200^0$C for 4 hours. Ethylene glycol acted both as solvent and reducing agent. After the reaction was complete, dark brown precipitate of magnetite was formed by partial reduction of Fe (III) salt. The nanoparticles were isolated by centrifugation following repeated wash with water and ethanol. The PAA is involved in surface functionalization and structural evolution of $Fe_3O_4$ nanospheres. Final shape of the nanoparticles is evolved through LaMer's method of fast super-saturated bust nucleation mechanism followed by aggregation and final growth to nanospheres [12].

Iron (Fe) nanorod was prepared by polyol-assisted solvothermal method. For the synthesis of Fe-nanorod, EG-hydrazine was used as reducing agents for the reduction of precursor solution. CTAB (Cetyl-trimethylammoniumbromide, 99%, Loba Chemie) was used as surfactant for directive growth of nanorods along the preferred plane. Fe (III) nitrate (0.02 M, Loba Chemie) and CTAB (0.05M) was dissolved to EG (30 mL) following rapid mechanical stirring. An aqueous solution of hydrazine hydrate (0.05M, Ranbaxy) was added to the precursor dropwise prior to transfer the resulting solution into a Teflon-lined stainless steel autoclave. The autoclave was kept for 7 hours inside a hot-air oven previously set to $200^0$C. The supernatant was removed by centrifugation, and the precipitate was vacuum-dried to yield the black powder. Since the relative yield of nanorod was less than 50%, repeated centrifugation of the precipitated black powder was necessary. The supernatant was rejected while heavier nanorods were collected at the bottom.

The MR fluids were made by dispersing proper amounts of $Fe_3O_4$ and Fe powder into silicone oil (viscosity 0.879 Pa.s at $25^0$C) through mechanical stirring and ultrasonication. Three MR suspensions (MRF1, MRF2 and MRF3) were prepared, with total dispersed phase concentrations maintained at 10 vol%. The amounts of spherical $Fe_3O_4$ were 10 vol%, 8 vol% and 5 vol% for MRF1, MRF2 and MRF3 respectively.

Particle morphology, size and shape were investigated by a field emission scanning electron microscope (Quanta FEG®, FEI). Crystal structures and phases of the powdered samples at room temperature were identified powder x-ray diffraction by using a PANalytical X'Pert PRO® diffractometer using monochromatic Cu-$K_\alpha$ radiation ($\lambda$=0.51418 nm). Room temperature magnetometric study of sample pellets (pressed powder) was performed on a Lakeshore Cryotonics® Inc. model 7400 VSM.

Field-dependent magnetorheological measurements were performed using a commercial rheometer (Anton Paar MCR Physica 501®) with magnetorheological attachment (MRD 170®) in strain-controlled mode and at room temperature. The parallel plate system with plate diameters of 20 mm was used for all measurements. A fixed plate-gap of 1 mm was maintained throughout the measurements. The magnetic field was generated vertically with respect to the direction of flow. Before any measurement, the sample was pre-sheared at 20 $s^{-1}$ for about 30 s. In oscillatory magnetosweep experiment, field was varied from 0 to 1.1 T under constant strain amplitude and frequency of 0.02% and 10 Hz, respectively. In amplitude sweep, an oscillatory strain ranging from 0.01% to 100% was applied to the sample under constant frequency of 10 Hz under constant uniform magnetic field of 0.33 T. Frequency sweep measurements were carried out under constant strain amplitude of 0.02% under the field of 0.33 T.

## 3. Results and discussions

*3.1. Morphology and magnetic characterization*

Morphological characterizations of Fe-nanorod and PAA-$Fe_3O_4$ performed in ESEM are shown in Fig.1 (inset). PAA-$Fe_3O_4$ nanoparticles are nearly spherical with high degree of surface roughness due to polymer-assisted growth in liquid polyol. While PAA-$Fe_3O_4$ nanospheres are mostly monodispersed, Fe-nanorods show high degree of polydispersity. Average diameter for PAA-$Fe_3O_4$ is found to be ~300 nm. Average length of Fe-nanorod is 800 nm with standard deviation of 100 nm. Cross-section of nanorod is measured as ~200 nm. It can be noted that relative yield of nanorod is less than 50% with significant presence of spherical nanoparticles.

X-ray diffraction patterns for $Fe_3O_4$ nanospheres and Fe-nanorods are shown in Fig. 1 (right). For magnetite, all peaks are assigned to spinel structure similar to $Fe_3O_4$ with crystallite size of 45 nm. The diffraction peaks at $30.4^0$, $35.88^0$, $37.5^0$, $43.3^0$, $53.1^0$, $57^0$, $62.73^0$ and $74.2^0$ are matched with standard magnetite XRD data (JCPDS No. 89-0691). Crystal structure of as-synthesized Fe-nanorods has also been confirmed by X-ray analysis. The peaks correspond to (110), (200) and (211) planes are characteristic to body centred cubic (bcc) structure of pure α-Fe (JCPDS No. 05-0696). Crystallite size calculated using Scherrer formula is 20.2 nm. It has been shown previously that the preferential growth in α-Fe nanorod occurs through the (200) plane [13]. Therefore, interaction of CTAB along the (200) plane may lead to epitaxial growth of nanorods.

Magnetic behavior of the nanostructures in powder form is investigated at room temperature (Fig. 2). The saturation magnetization ($M_s$) and coercivity ($H_c$) for $Fe_3O_4$ are measured to be 61.7 emu/g and 51 Oe whereas for α-Fe, $M_s$ and $H_c$ are 120.5 emu/g and 541 Oe, respectively. $M_s$ of $Fe_3O_4$ is smaller in magnitude than bulk value for magnetite, but significantly higher than that reported earlier. However, reduction in saturation magnetization in nanoparticles compared to that of bulk value can originate due to number of factors, including association of polymer in composite nanostructures, surface spin canting and presence of magnetic dead layers on surface [14]. Room temperature magnetization curve for Fe-nanorod showed ferromagnetic behavior with high $M_s$, characteristic to α-Fe. However, observed $M_s$ is still much smaller than that of bulk saturation magnetization of pure Fe (218 emu/g at 300K). One of the reasons is the association of surfactant CTAB in the growth of nanorods.

*3.2. Magnetorheological characterization*

Fig. 3 displays the dependence of the storage ($G'$) and loss moduli ($G''$) for MRFs as a function of deformation amplitude at a fixed frequency (10 Hz) and magnetic field strength of 0.33T. The loss factor (tanδ) is also plotted in Fig. 4 inset. The MR samples are pre-treated with homogeneous magnetic field for 3 minutes. This is to allow the samples to form equilibrium field-induced microstructures before the strain sweep would take place. In presence of magnetic field, MRFs behave desirably with steady decline in storage modulus as a function of strain amplitude. Characteristic to field-induced $G'$ for aggregated microstructures, short LVE region (~0.002 strain %) is reported for all the suspensions. The trend in initial value of $G'$ follows the order: MRF1<MRF2<MRF3. This is expected as with higher substitution of Fe-nanorod, magnetic and shape anisotropy ensures stronger structuration. The magnitude of magnetorheological Payne effect ($G_0' - G_\infty'$) also increases with the increasing Fe-nanorod content. At low Fe loading, the observed variation in the amplitude of the Payne effect is weak. But as the vol% of Fe increases, significant and pronounced variation is observed. This is principally due to the breakdown of aggregated networks at higher strains [7, 8]. With increasing strain, unperturbed chains of magnetic particles start to crumble. The strength of interparticle network is a function of deformation as the dipole-dipole interaction energy between particles decreases with increasing strain. Upon increasing the strain further, steady decline in $G'$ is reported for all MRFs. This can be explained in terms of magnetorheological Payne effect, where distance between particles in the aggregates is increased with applied strain, implying a drop in magnetic dipole-dipole interaction energy [8]. In presence of field, beyond the critical strain corresponds to LVR, the distribution of interparticle distance is no more even, this is to say that response of magnetized clusters will be completely different than what it was in LVR. The destruction of clusters and their reformation and redistribution in presence of field at higher deformation gives rise to nonlinear decay in storage modulus.

The trend in loss moduli (Fig. 3 inset) illustrates three distinct regions: at very low strain (up to 0.03%), G″ is practically independent; after it increases with increasing strain to form a maxima at γ~1%. Following a cross-over point with G′, a long decreasing tail is observed. The short linear region basically coincides with LVR, following an increase in magnitude of G″ till it reaches a maximum. Increase in the loss moduli can be explained in terms of dissipation of energy for continuous rupture of the magnetic coupling between magnetized particles with increasing γ. The subsequent modulus decrease is evident when the microstructure is already started collapsing at higher γ and the interparticle interactions decrease due to growing distances between particles. Beyond the cross-over strain ($γ_{cr}$), flow is almost laminar and therefore, no significant change in G″ is observed. The loss factor curves for the MRFs under similar conditions reveal a few interesting aspects of the field-induced dynamic rheological property. At small initial γ, tanδ is independent of strain. Afterwards, a rapid increment is observed with continuous increase in periodic deformation. General nature of loss tangent curves are similar for all samples, showing sharp increase at γ = 0.1-1% and then become independent of strain. Initial increase represents the viscous energy loss due to stronger interparticle attraction. Beyond crossover point at very high strain, it becomes independent, implying a permanent rupture of microstructures. An interesting observation is the similarities in rheological response for the MRFs. Despite the difference in the amount of anisotropic Fe-nanorod content, MRFs behave nearly identical except at the initial values of moduli. A similar response indicates a similar on-field structural arrangement. Despite the fact that MRFs contain different isotropic and anisotropic magnetic dispersed phases, on-field rheological behavior is dynamically identical. The higher initial values of dynamic moduli ($G_0′$ and $G_0″$) indicate stronger structures owing to increased shape and magnetic anisotropy.

It is seen that the dynamic breakdown and reformation of fractal network comprised of nanospheres and nanorods with increasing deformation gives rise to magnetorheological Payne effect, similar to that of carbon-filler containing rubber matrix [7]. The hydrodynamic interaction between filler networks in carbon-rubber filler gives rise to stress relaxation phenomenon [6]. Similarly the interplay between magnetostatic interaction of aggregated networks and hydrodynamic forces arising from mechanical strain gives rise to magnetorheological Payne effect. A few mathematical models were derived subsequently in order to explain the Payne effect, of which the one proposed by Kraus comes first [15]. Kraus constructed a phenomenological model which has often been applied for the fitting of experimental data. The model is based on the assumption that the filler aggregates break and recombine under different rates, which depend on amplitude of deformation and on rate constants. In this model, however, the interaction between filler network and matrix polymer was ignored. Since there is little or no chemical interaction between magnetic particles and carrier fluid in a typical MRF, the magnetostatic interaction clearly dominates over all other interactions. The result says that the excess storage modulus ($\frac{G' - G'_\infty}{G'_0 - G'_\infty}$) with increasing deformation amplitude has the following characteristic form:

$$\frac{G' - G'_\infty}{G'_0 - G'_\infty} = \frac{1}{1 + \left(\frac{\gamma}{k}\right)^{2m}}$$

… (1)

Where, k and m are fitting parameters; since Kraus model equation is based on some arbitrary assumptions, the fitting parameters do not possess any direct obvious physical significance [16]. The quantification of Payne effect was based on agglomeration and de-agglomeration of fractal networks, the parameter k represents a rate constant of structural aggregation and subsequent rupture. It is expressed as,

$$k = \left(\frac{k_A}{k_B}\right)^{\frac{1}{2m}}$$

… (2)

Where, $k_A$ and $k_B$ are rate constants for agglomeration and de-agglomeration, respectively. The parameter m is considered to be universal for carbon filler-rubber composite materials. Both the parameters are empirical and can't be evaluated from direct measurements. The fitting of G′ versus γ plots with Kraus model is shown in Fig. 4. The fitting parameters are listed in Table 1.

**Table 1** fitting parameters of Kraus fit for G′ versus strain% (γ) for MRFs

| Fitting parameter/samples | MRF1 | MRF2 | MRF3 |
|---|---|---|---|
| k | 0.181 | 0.212 | 0.232 |
| m | 0.75 | 0.76 | 0.77 |

One can see that the fittings are remarkably good. However, the values of parameters are not constant, although the scatter in k and m values is quite narrow. An increasing trend in k can be qualitatively explained by the structural integrity in samples with increasing Fe-nanorod contents. It is obvious that rate constant for agglomeration is higher in highly anisotropic MRF, compared to the isotropic spherical nanoparticle-containing MRF1. For m values, slight increment in magnitude is reported for MRFs with increasing structural and magnetic anisotropy.

It is important to note that Payne effect originating from non-magnetic filler contents in polymer matrix is different from that of the magnetorheological Payne effect. In contrast to matrix-filler interaction, magnetorheological fluids demonstrate both short and long range magnetostatic interaction. Even when the strain is sufficiently high, there is long-range interaction between particles. This is probably the reason that our reported values of m are significantly higher than that reported by previous authors [7, 8].

Magnetosweep studies under oscillation are performed at very low constant strain amplitude of 0.02%. Unlike rotational rheometry, oscillatory tests at sufficiently low deformation do not cause irreversible rupture of aggregates. As observed in amplitude sweep test, MRFs possess very narrow LVR region. For magnetosweep study, a constant angular frequency of 10 Hz was applied while increasing magnetic field continuously. The plots of G′ of MRFs as a function of field are shown in Fig. 4 inset. As expected, MRF3 with highest nanorod concentration exhibits highest percent increase in G′ when field sweeps from 0 to 1 T, followed by MRF2 and MRF1. The overall behavior can be generalized in terms of three distinct regions: (I) in the first zone, increment in G′ value is very small with increasing field. At low field, conglomerates of magnetic structures are more evenly distributed, giving rise to less steeper response. In region (II) of intermediate field range, a linear increase is reported owing to the formation of more robust aggregates. This region illustrates the formation of quasi-gel or solid-like structures from liquid-like state. Beyond the intermediate field range, region (III) shows a somewhat hard gel like behavior where the sample becomes almost resistant to flow. This happens because particles and aggregates are fully aligned along the field lines.

Frequency-dependent oscillatory measurements are recorded under very low constant amplitude of deformation (0.02%) and the frequency was varied in the range of 0.1-100 Hz. Frequency-dependence of storage and loss modulus measured in presence of magnetic field (0.33 T) is shown in Fig. 5. Under small angle oscillatory strain (SAOS), MRFs shows no significant critical behavior. As expected, G′ of the anisotropic MRFs display stronger frequency dependence owing to stronger aggregate formation under small strain. The slope of G′ versus frequency curves shows regular increase for the samples in this order: MRF3>MRF2>MRF1. For SAOS, no significant change in rheology is expected except the redistribution of microstructural aggregates under increasing frequency. In anisotropic samples (MRF2 and MRF3), some pre-ordering of the particle aggregates in absence of magnetic field results in strengthening of the field induced chain-structures, therefore, giving rise to an increase in the mechanical modulus (G′). However, behavior of los modulus is difficult to explain. Particularly, the origin of hump in G″ curves in the frequency region 2-10 Hz is unclear. The viscous dissipation of energy is associated to hydrodynamic effect which can be attributed to internal movement of aggregates with increasing frequency [17]. But it is not yet clear which mechanism influences the magnitude and frequency-dependent trend of G″.

## 4. Conclusion

In this paper, we have described typical one-pot synthesis of magnetite nanospheres by thermal decomposition of Fe (III) acetylacetonate precursor in polyol and polyacrylic acid. Fe-nanorods were prepared by a novel polyol-hydrazine-CTAB assisted reduction of Fe (III) nitrate. Morphology and magnetic characterization of the nanoparticles were performed by ESEM, XRD and VSM studies. We performed detailed non-linear magnetorheological properties of three MR fluids (10 vol%) containing isotropic $Fe_3O_4$ and anisotropic Fe-nanorods under both small (SAOS) and large amplitude oscillatory deformation (LAOS). The MR suspensions demonstrated rapid stress relaxation under increasing deformation and uniform magnetic field beyond linear viscoelastic region (LVR), which has not been studied in-depth in conventional MR fluids. The stress relaxation under increasing strain is termed as Payne effect. We have also shown that stress softening was more pronounced for MR fluids with higher anisotropic contents (MRF2 and MRF3), in contrast to isotropic MR fluid. The threshold strains for LVR to non-linear region transition for MRF2 and MRF3 were smaller in magnitude than that of isotropic spherical nanoparticle-containing fluid (MRF1). The fixed strain magnetosweep experiment in oscillatory flow displayed three distinct regions in field-dependent curves. This is characteristic to magnetorheological fluids and can be interpreted in terms of field-induced structural inhomogeneity within the fluid.

## Acknowledgments

One of the authors, IA thanks CSIR, India for the award of Senior Research Fellowship (SRF).

## References

[1] G. Bossis, S. Lacis, A. Meunier, O. Volkova, J. Magn. Magn. Mater. 252 (2002) 224–228.

[2] C. Galindo-Gonzalez, M. T. Lopez-Lopez, J. D. G. Duran, J. App. Phys. 112 (2012) 043917.

[3] W. Schutt, J. Teller, U. Hafeli, M. Zborowski (Eds.), Scientific and Clinical Applications of Magnetic Carriers, Plenum, New York, 1997.

[4] A. Gomez-Ramirez, M. T. Lopez-Lopez, J. D. G. Duran, F. Gonzalez-Caballero, Soft Matter 5 (2009) 3888–3895.

[5] J. de Vicente, J. P. Segovia-Gutiérrez, E. Andablo-Reyes, F. Vereda, R. Hidalgo-Álvarez, J. App. Phys. 131 (2009) 194902.

[6] A. R. Payne, J. Appl. Polym. Sci. 6 (1962) 57–63.

[7] V. V. Sorokin, E. Ecker, G. V. Stepanov, M. Shamonin, G. J. Monkman, E.Y. Kramarenko, A. R. Khokhlov, Soft Matter 10 (2014) 8765–8776.

[8] H. An, S. J. Picken, E. Mendes, Polymer 53 (2012) 4164e4170.

[9] B. J. Park, M. S. Kim, H. J. Choi, Materials Letters 63 (2009) 2178–2180.

[10] G. T. Ngatu, N. M. Wereley, J. O. Karli, R. C. Bell, Smart Mater. Struct. 17 (2008) 045022.

[11] Y. D. Liu, H. J. Choi, J. App. Phys. 115 (2014) 17B529.

[12] V. K. LaMer, R. H. Dinegar, J. Am. Chem. Soc. 72 (1950) 4847.

[13] T. Fujita, M. Chen, X. Wang, B. Xu, K. Inoke, K. Yamamoto, J. App. Phys. 101 (2007) 014323.

[14] I. Arief, P. K. Mukhopadhyay, J. Magn. Magn. Mater. 372 (2014) 214–223.

[15] G. Kraus, J. Appl. Polym. Symp. 39 (1984) 75.

[16] G. Hubery, T. A. Vilgisy, G. Heinrich, J. Phys.: Condens. Matter 8 (1996) L409–L412.

[17] K. Shah, J. Oh, S. Choi, R. V. Upadhyay, J. Appl. Phys. 114 (2013) 213904.

**Figures with captions:**

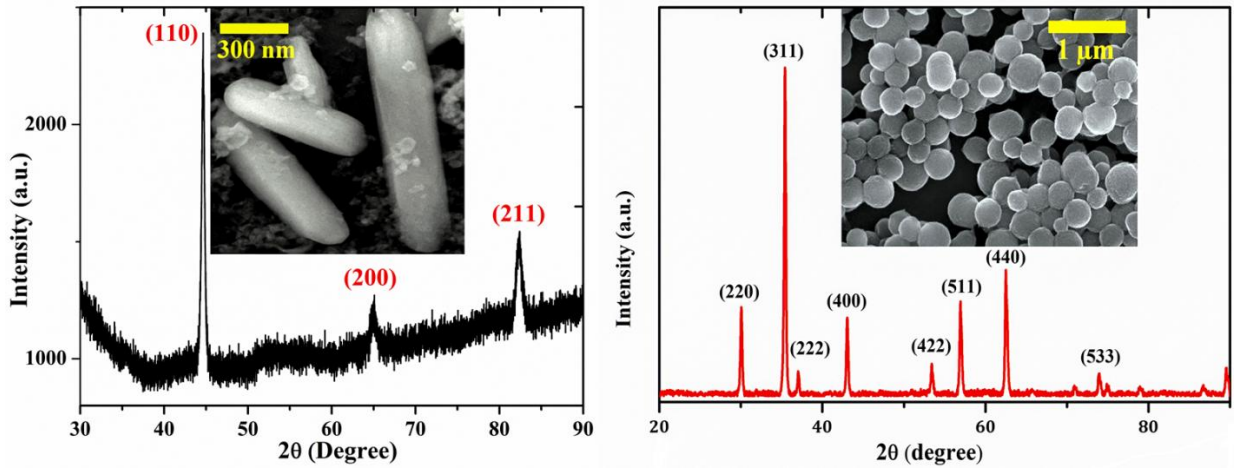

**Fig. 1.** X-ray diffraction patterns of Fe-nanorod (left) and PAA-Fe$_3$O$_4$ nanospheres (right). ESEM images of as-prepared Fe-nanorod (left, inset) and PAA-Fe$_3$O$_4$ nanospheres (right, inset) were also shown.

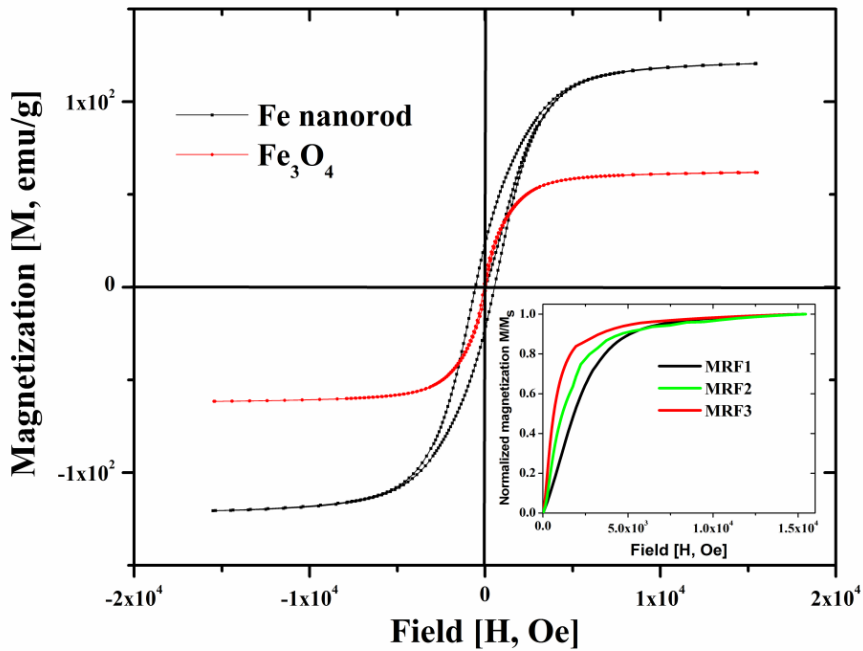

**Fig. 2.** Room temperature magnetization versus field hysteresis curves (M-H) were shown for Fe-nanorod (black symbols) and Fe$_3$O$_4$ nanospheres (red symbols), respectively. Inset, normalized magnetization (M/M$_s$) of the magnetorheological fluids were plotted as a function of magnetic fields.

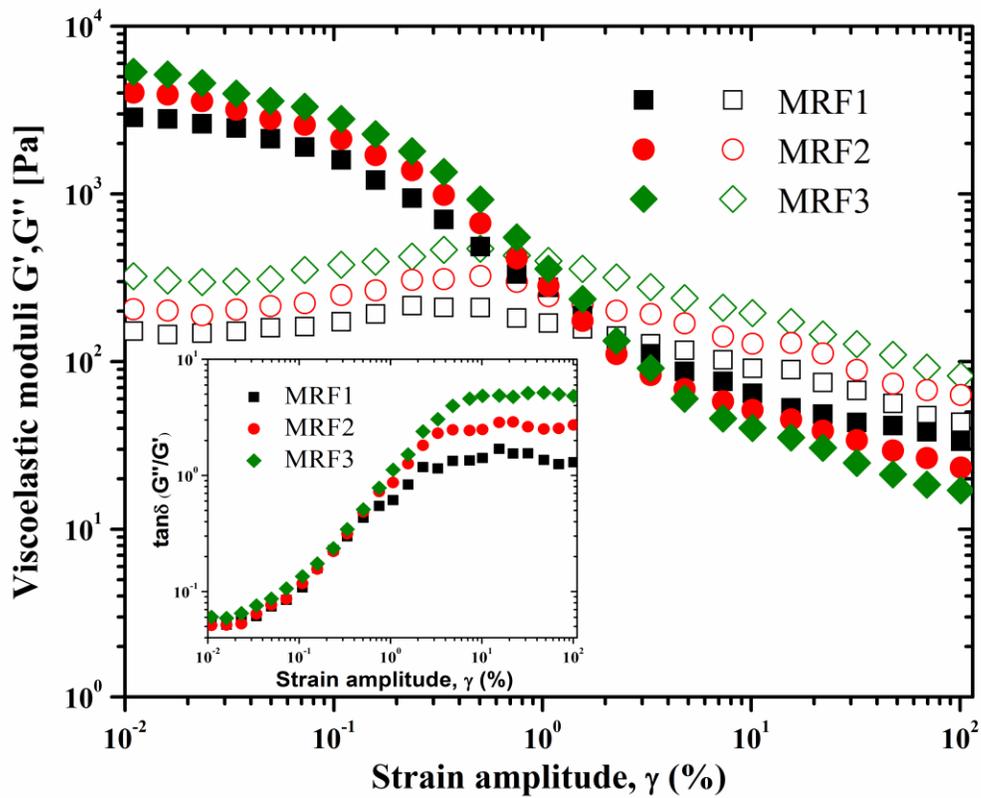

**Fig. 3.** Amplitude sweep oscillatory rheology for MRF1 (black symbols), MRF2 (red symbols) and MRF3 (green symbols) under constant frequency of 10 Hz and constant magnetic field 0.33T. Storage (G′) and loss moduli (G″) represented by closed and open symbols, respectively. Inset, loss factor (G″/G′) was plotted against strain amplitude.

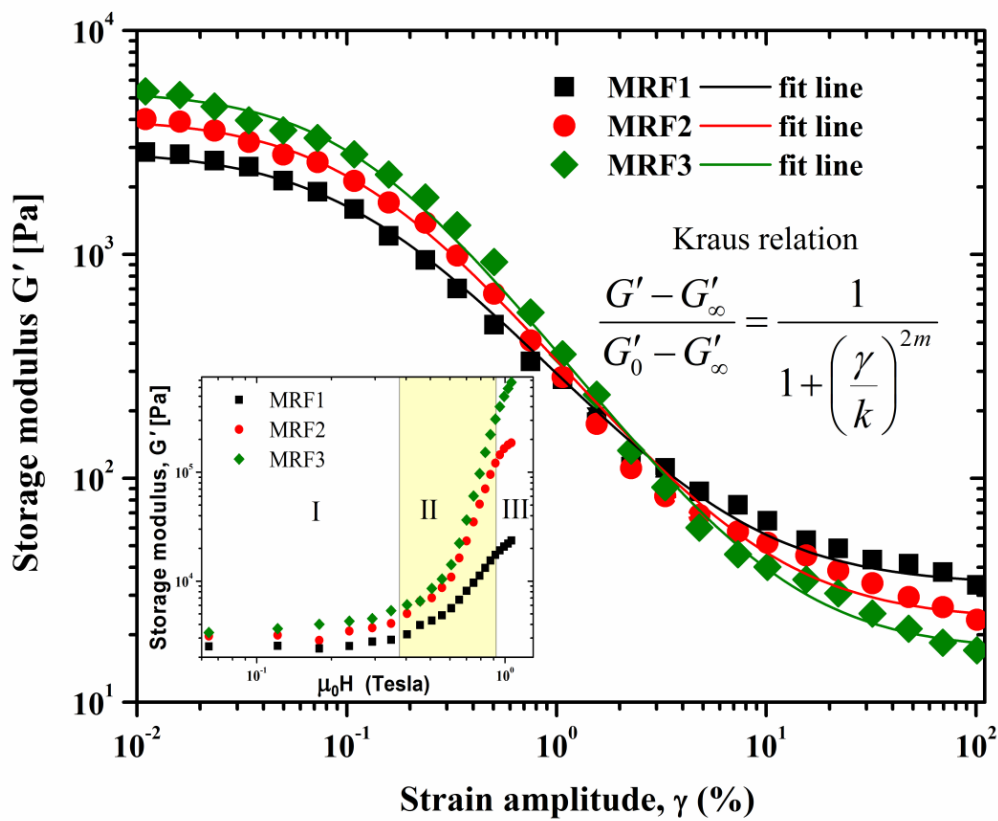

**Fig. 4.** Storage modulus (G′) is plotted as a function of strain in amplitude sweep oscillatory measurement under constant frequency of 10 Hz and constant magnetic field 0.33T. Fit lines represent Kraus phenomenological equation relating the change in G′ with strain (γ); $k$ and $m$ are fitting parameters. Inset, magnetic field sweep of storage modulus (G′) at constant strain amplitude of 0.02%.

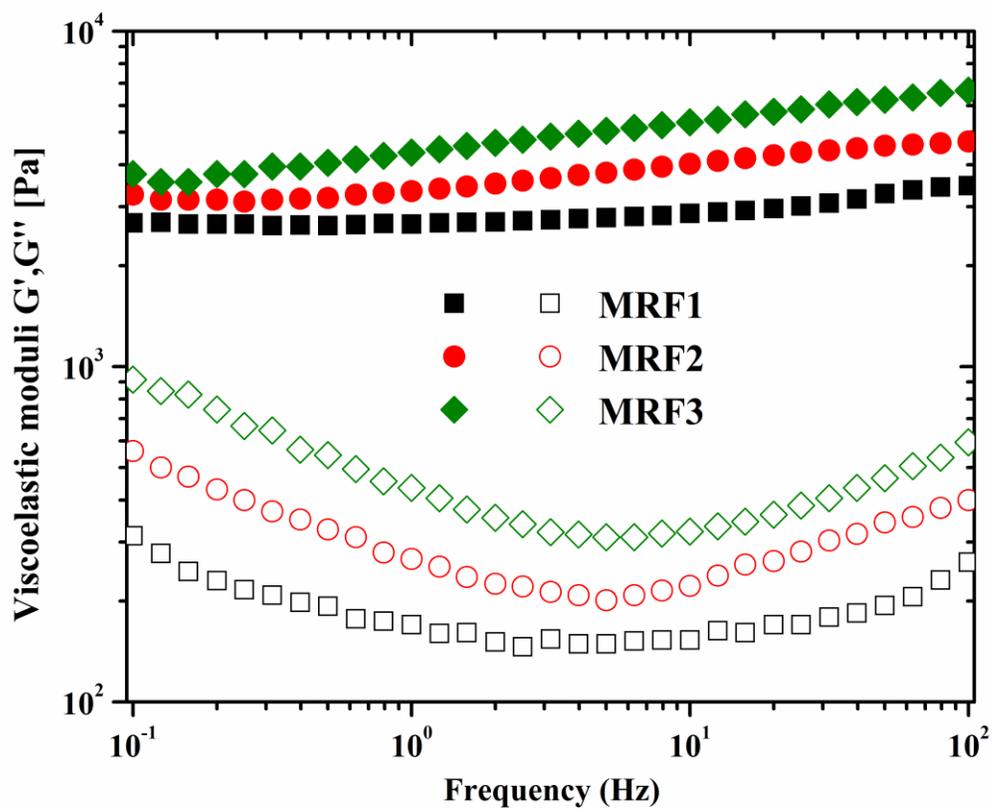

**Fig. 5.** Frequency sweep oscillatory rheological curves were shown for MRF1 (black squares), MRF2 (red circles) and MRF3 (green diamonds) under constant strain amplitude (γ) of 0.02% and constant magnetic field of 0.33T. Storage (G′) and loss moduli (G″) represented by closed and open symbols, respectively.